\documentclass[aps,prl,reprint,superscriptaddress]{revtex4-2} 
\usepackage{amsmath,amssymb}
\usepackage{graphicx}
\usepackage{listings}
\usepackage{siunitx}
\usepackage{comment}
\usepackage{xcolor}
\begin{document}
\title{Energy dissipation at the atomic scale explains how fracture energy depends on crack velocity in silica glass}
\author{Marthe Grønlie Guren}
\affiliation{The Njord Centre, Departments of Geosciences and Physics, University of Oslo}
\affiliation{Expert Analytics, Oslo, Norway}
\author{Sigbjørn Løland Bore}
\affiliation{The Hylleraas Centre, Department of Chemistry, University of Oslo}
\author{Fran\c cois Renard}
\affiliation{The Njord Centre, Departments of Geosciences and Physics, University of Oslo}
\affiliation{Univ. Grenoble Alpes, Grenoble INP, Univ. Savoie Mont Blanc, CNRS, IRD, Univ. Gustave Eiffel, ISTerre, Grenoble, France}
\author{Henrik Andersen Sveinsson}
\affiliation{The Njord Centre, Departments of Geosciences and Physics, University of Oslo}
\date{\today}
\begin{abstract}
The fracture energy of brittle materials rises with crack velocity, and this effect is typically attributed to surface roughening from path instabilities. Here we show, using molecular dynamics simulations of silica glass with a first-principles machine learned interatomic potential, that the structural fracture energy rises by up to 33\% already below the branching threshold, showing that fracture energy is not a constant material property. This rise in fracture energy is roughly equally partitioned between an increase in the intrinsic surface energy density and nanoscale roughening that increases the real fracture surface area. Results demonstrate that dynamic fracture in silica glass increases the fracture energy not merely by creating more apparent surface, but also by creating a fundamentally different surface at the nanoscale.
\end{abstract}
\maketitle
\section{Introduction}
Crack propagation in silica glass requires a fracture energy that far exceeds the thermodynamic cost of bond rupture. Even under quasi-static conditions, experiments show a significantly higher fracture energy than predictions based purely on surface energy and linear elastic fracture mechanics \cite{wiederhorn1969fracture}. This discrepancy widens dramatically as the crack velocity increases, creating a surplus of energy that drives complex dissipation phenomena. In brittle materials like silica glass, a significant portion of this excess energy is dissipated as heat \cite{weichertHeatGenerationTip1978a}, generating localized temperature spikes of several thousand Kelvin at the crack tip. A central question remains: How much is stored as post-mortem surface energy, and in what way? Do the extreme temperatures at the crack tip fundamentally alter the fracture surface itself?

As a crack accelerates, dynamic instabilities can cause the tip to oscillate or branch \cite{finebergInstabilityDynamicFracture1991, buehler2006dynamical}, creating additional surface area through roughening. \citet{sharonEnergyDissipationDynamic1996} studied fracturing of Poly(methyl methacrylate)--PMMA--and argued that the velocity-dependent fracture energy is a dynamic effect of microbranching. Beyond a critical velocity, the rate of surface creation becomes proportional to the energy flux, suggesting $\gamma_{\rm s}$, the energy cost to create a unit of relaxed surface area, remains a constant material property. Yet they also observed a 30~\% increase in fracture energy below the onset of microbranching, which they attributed to unresolved dissipation processes \cite{sharonEnergyDissipationDynamic1996}. Whether this pre-branching increase arises from a change in surface area (quantity), through nanoscale roughening invisible to their measurements, or a change in the energy density of the created surface, driven by the extreme conditions at the crack tip (quality), has remained an open question.

Molecular dynamics simulations offer the necessary spatio-temporal resolution to resolve nanoscale processes, but previous studies of fracture in silica have relied on empirical potentials. These potentials have successfully reproduced structural properties at low strains \cite{vashishtaInteractionPotentialSiO21990} and demonstrated a fracture process zone of around \SI{10}{\nano\meter} \cite{rountree2010fracture}, but the empirical potentials have failed to capture the details of high-energy bond-breaking events, leading for example to underestimated branching speeds during dynamic fracture \cite{guren2022nanoscale}.

Modeling from first principles offers an unbiased description of bond-breaking and bond-forming at the nanoscale. Machine learned interatomic potentials (MLIPs) \cite{behlerGeneralizedNeuralNetworkRepresentation2007}, trained to predict the solution of the many-body electronic structure problem at a fraction of the computational cost of full Density Functional Theory simulations, now make it possible to combine this accuracy with the system sizes needed to study crack propagation at realistic scales \cite{zengDeePMDkitV2Software2023, NEURIPS2022_4a36c3c5, musaelianLearningLocalEquivariant2023}. For silica glass, several MLIPs have been trained on Density Functional Theory data using various model architectures \cite{erhardMachinelearnedInteratomicPotential2022, kobayashiMachineLearningMolecular2023,novikov2019improving, balyakin2020deep}, but none have previously been applied at the scale required to investigate fracture process zones and crack instabilities during dynamic fracture.

Here, we use extensive first-principles molecular dynamics simulations to resolve the origin of velocity-dependent fracture energy in silica glass. We find that even below the branching threshold at 0.72 of the Rayleigh velocity of the solid, $v_R$, the structural fracture energy measured post-mortem rises by up to 33~\%. This increase is driven roughly equally by nanoscale roughening--which increases the real fracture surface area--and an increase in the intrinsic surface energy density. Crucially, the nanoscale roughening would be invisible to standard post-mortem analysis, meaning that typical fracture experiments would register all of the increase as an increase in surface energy density. Our results demonstrate that dynamic fracture in silica glass increases the fracture energy both by creating more surface area at the nanoscale, and by increasing the intrinsic energy density of that surface.
\begin{figure}
    \centering
    \includegraphics[width=\linewidth]{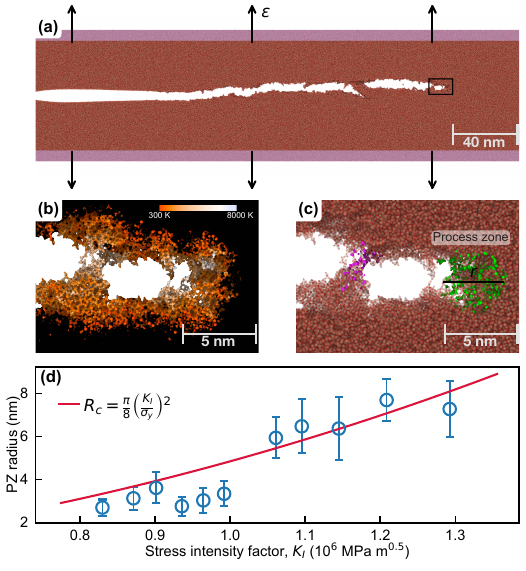}
    \caption{Simulation setup overview and crack process zone determination (a) The three-dimensional simulation setup with a running crack, and the loading direction indicated with arrows. Shaded regions at the top and bottom indicate frozen atoms where the tensile loading $\varepsilon$ is applied. The crack propagates to the right from a pre-existing notch at the left boundary. The clean crack surface of the leftmost \SI{75}{\nano\meter} of the system shows the location of the initial notch. The small black rectangle marks the domain that is magnified in panels (b) and (c). (b) Per-atom kinetic temperature field at the crack tip during propagation at $\sim$\SI{2.4}{\kilo\meter\per\second}, showing local temperatures exceeding \SI{8000}{\kelvin} within the process zone. (c) Under-coordinated atoms at the crack tip. The largest under-coordinated cluster (green) defines the process zone (PZ) ahead of the crack tip, while the second-largest cluster (pink) trails behind it. (d) Process zone radius as a function of the stress intensity factor. The solid red line is the Dugdale-Barenblatt prediction with $\sigma_y= \SI{10}{\giga\pascal}$.}
    \label{fig:setup}
\end{figure}
\section{Simulations}
We model a pure silica glass (fused silica) from first principles, using a machine learned interatomic potential (MLIP) representing a Density Functional Theory (DFT) energy surface at the r$^2$SCAN \cite{furnessAccurateNumericallyEfficient2020} level of theory. We use an Allegro \cite{musaelianLearningLocalEquivariant2023} MLIP that was trained on 1982 silica structures, mostly molten or glassy and under various states of strain. The training data and the interaction potential parameters are provided as Supplementary Data (zenodo). See extended methods for details about the training of the MLIP and the settings of the underlying r$^2$SCAN calculations.

The simulation setup consists of a silica glass in a thin strip geometry with dimensions $300 \times 8.5 \times \SI{75}{\nano\meter}^3$, with a \SI{75}{\nano\meter} long pre-crack notch running in the $x$ direction (Fig. \ref{fig:setup}a). The simulation setup contains $12.3$ million atoms. The systems were stabilized at \SI{300}{\kelvin} and \SI{1}{\bar} and then rapidly, over \SI{50}{\pico\second}, subjected to an extensional displacement along the $z$-direction, perpendicular to the pre-crack. We ran 14 simulations with varying displacement rates leading to initial tensile stresses in the range \SIrange{3.9}{7}{\giga\pascal}. The simulations were then run for \SIrange{0.3}{1}{\nano\second}, to observe crack propagation. We observed fast crack propagation in simulations with stresses exceeding \SI{4.02}{\giga\pascal}, and we observed signs of slow crack growth at \SI{3.94}{\giga\pascal}. We therefore set the critical stress necessary to propagate a crack in our thin strip geometry to be \SI{3.94}{\giga\pascal}. Using the formula for stress intensity, $K_{I}$, in a thin strip \cite{buehlerAtomisticModelingMaterials2008}, $K_{I} = \sigma\sqrt{\xi/2}$, where $\xi$ is the strip height, we obtain the critical intensity factor, $K_{Ic} = \SI{0.763}{\mega\pascal\sqrt{\meter}}$ for our model fused silica, which is in excellent agreement with experimental data ($\SI{0.798}{\mega\pascal\sqrt{\meter}}$ \cite{wiederhorn1969fracture}).
\begin{figure}
    \centering
    \includegraphics[width=\linewidth]{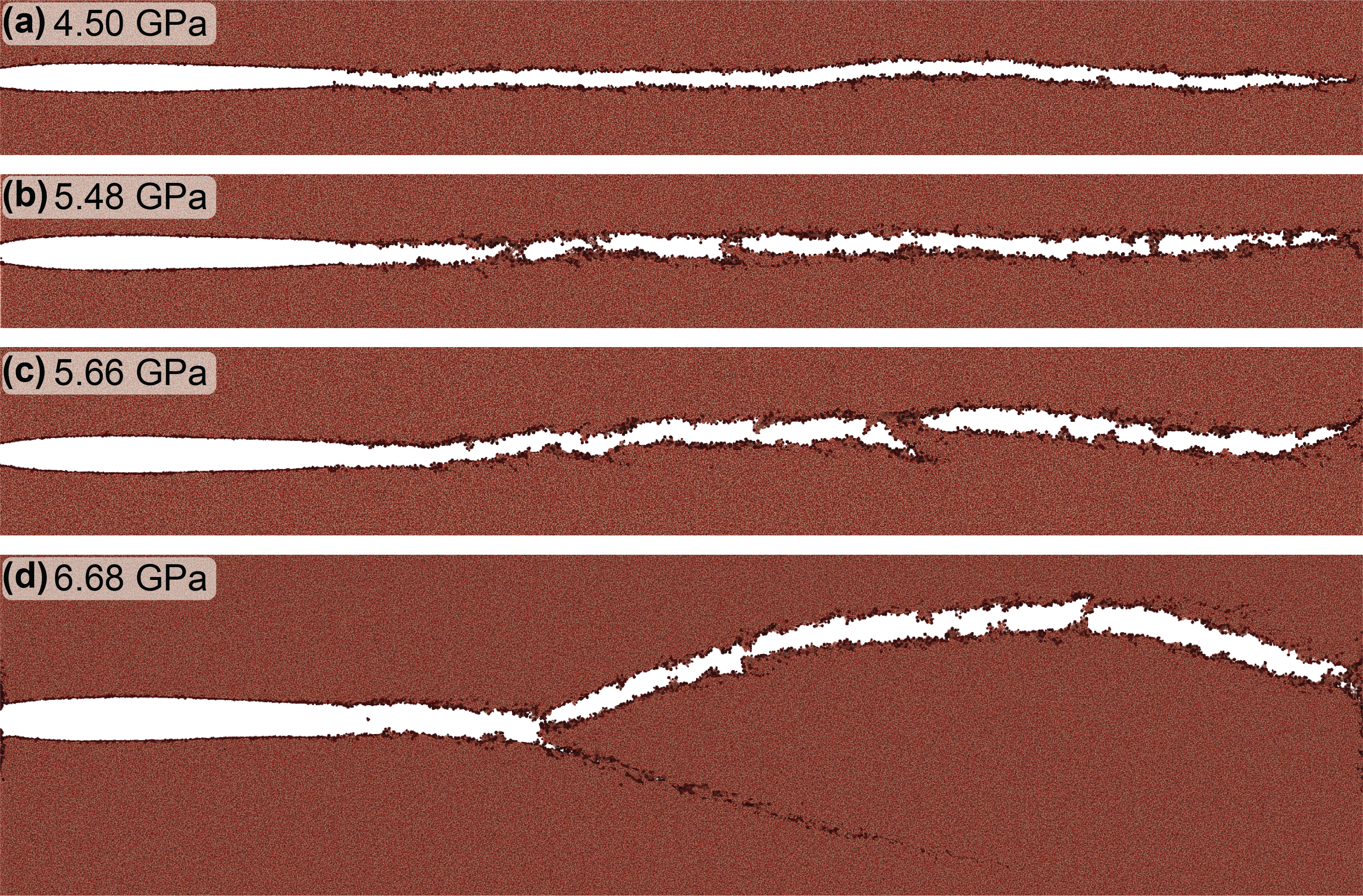}
    \caption{Post-mortem fracture morphology for four representative simulations at increasing initial tensile stress. Under-coordinated atoms (dark red) mark the crack surfaces and the surrounding damage zone. With increasing stress, the damage zone thickens and the crack transitions from planar propagation to branching.}
    \label{fig:morphology}
\end{figure}
\subsection{Fracture morphology and fracture process zone}
We classify each crack resulting from a crack simulation as either a planar crack or a branching crack. Fig. \ref{fig:morphology} displays snapshots from four different simulations with increasing initial loading stresses, taken after crack propagation. In these snapshots, we show under-coordinated atoms in dark red (i.e., coordination lower than two for oxygen and lower than four for silicon), since these atoms represent the crack surface and the damaged volume around the crack. At low initial tensile stress ($\sigma = \SI{4.50}{\giga\pascal}$, Fig. \ref{fig:morphology}a), cracks are straight and the damaged area is thin. With increasing initial tensile stress, there is more damage near the crack surface, which can be observed by a thicker layer of undercoordinated particles surrounding the fracture (Fig. \ref{fig:morphology}b). At higher stresses (Fig. \ref{fig:morphology}c--d) we observe crack branching. Branching is observed when the fracture splits at the crack tip, and two fractures propagate simultaneously. For example, at $\sigma = \SI{5.66}{\giga\pascal}$ (Fig. \ref{fig:morphology}c), the crack makes several branching attempts, while at $\sigma = \SI{6.68}{\giga\pascal}$ (Fig. \ref{fig:morphology}d), a second branch propagates more than ${\sim}\SI{100}{\nano\meter}$.

The size of the process zone in silica glass has previously been measured to be \SI{10}{\nano\meter} in molecular dynamics simulations \cite{rountree2010fracture} and \SIrange{15}{35}{\nano\meter} with atomic force microscopy \cite{rountree2020silica}. We observe that the process zone's dimension increases with increasing stress intensity factor ($K_I$), and especially when branching is occurring (Fig. \ref{fig:setup}d). The sizes of process zones in our simulations are $r=\SI{3.1 \pm 0.3}{\nano\meter}$ for straight cracks and $r=\SI{6.8 \pm 0.6}{\nano\meter}$ for branching cracks, corresponding to  a process zone of $229 \pm 55$ atoms and $1046 \pm 252$ atoms, respectively in our particular simulation setup. Note that this increase of the process zone dimension is not due to the formation of two separate branches of the crack. We measure only the largest process zone cluster, so only the process zone of the main branch will contribute to the process zone estimate.

We compare these process zone dimensions to the prediction of the Dugdale--Barenblatt expression \cite{dugdale1960yielding,barenblatt1962mathematical}, $R_C = \frac{\pi}{8} \left(\frac{K_I}{\sigma_y}\right)^2$ shown as a solid line in Fig. \ref{fig:setup}d. We have inserted our simulation values for the thin strip $K_I$ and a literature estimate for the yield stress $\sigma_y$ (here set to \SI{10}{\giga\pascal}). The Dugdale--Barenblatt expression is in relatively good agreement with our results, although our simulations point to more of a step increase than a continuous quadratic increase when cracks evolve from planar to branching.
\subsection{Crack velocities and instabilities}
\begin{figure}
    \centering
    \includegraphics[width=\linewidth]{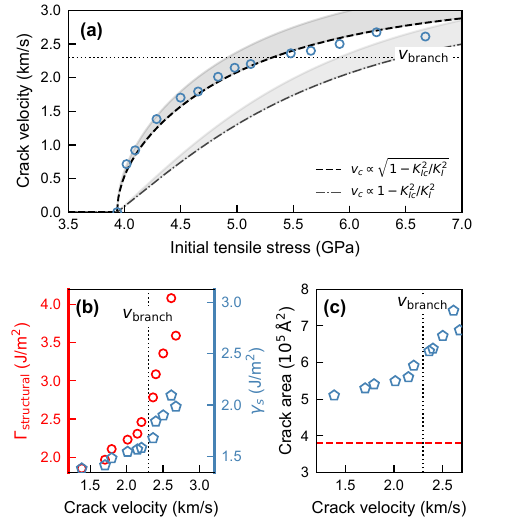}
    \caption{Crack velocities and fracture surface energies. (a) Crack tip velocity $v_c$ as a function of initial tensile stress. The dashed-dotted line is the standard Freund model (Eq.~\ref{eq:vc_classic}), and the dashed line is the square-root relationship (Eq.~\ref{eq:vc_sqrt}). The gray shaded area shows the effect of increasing Young's modulus from 68.9 to \SI{90}{\giga\pascal}, to illustrate the magnitude of the effect of hyperelasticity. The horizontal dotted line marks the branching velocity $v_{\text{branch}} \approx 0.72 v_R$. (b) Post-mortem fracture energy decomposition as a function of crack velocity. Both quantities measure the structural energy increase from intact to cracked state, after cooling down, excluding dissipated heat. Red circles (left axis): structural fracture energy $\Gamma_{\mathrm{structural}}$ (Eq.~\ref{eq:gamma_effective}), normalized by the projected crack area. Blue circles (right axis): Intrinsic surface energy density $\gamma_s$ (Eq.~\ref{eq:gamma_real}), normalized by the real (rough) fracture surface area. The vertical dashed line marks $v_\mathrm{branch}$. (c) Real fracture surface area $A_{\text{real}}$  as a function of velocity. The dashed red line indicates twice the constant projected area of a perfectly planar crack.}
    \label{fig:velocity}
\end{figure}

Our simulations show an increase in the crack velocity with increasing applied tensile stress. The exact observed relationship in our simulations is shown in Fig.~\ref{fig:velocity}a. The crack speed increases with stress up to a plateau at $0.72\,v_R$, beyond which the cracks branch. We now compare these results to analytical velocity--stress relationships. 
The standard Freund model \cite{freund1998dynamic} relates crack speed, $v_c$, to stress intensity via the Rayleigh wave speed $v_R$:
  \begin{equation}
      v_c = v_R\left[1-\left(\frac{K_{Ic}}{K_I}\right)^2\right].
      \label{eq:vc_classic}
  \end{equation}
Our simulation data are instead well described by a square-root form (Fig.~\ref{fig:velocity}a):
  \begin{equation}
      v_c = v_R\sqrt{1-\left(\frac{K_{Ic}}{K_I}\right)^2},
      \label{eq:vc_sqrt}
  \end{equation}
where the only fit parameter is $K_{Ic}$. The remaining quantities ($E=\SI{68.9}{\giga\pascal}$, $\nu=0.183$, $\rho_0=\SI{2.17}{\gram\per\centi\meter\cubed}$) are obtained from an independent simulation we ran to measure the elastic properties of our model silica glass. The fit to Eq. \ref{eq:vc_sqrt} is plotted along with the simulation data in Fig. \ref{fig:velocity}a, with a gray shaded area showing the effect of increasing Young's modulus up to \SI{90}{\giga\pascal}. Young's modulus varies with strain (Fig. \ref{fig:E_nu}), but this effect is much smaller than the difference between Eqs \ref{eq:vc_classic} and \ref{eq:vc_sqrt}. The fit to Eq. \ref{eq:vc_sqrt} holds well up to the branching speed, but above it the measured velocities fall below the prediction. The departure from the standard Freund model (Eq.~\ref{eq:vc_classic}) can probably be attributed to the thin strip geometry, where the fixed-displacement boundaries impose a constant energy release rate, unlike the semi-infinite geometry where elastodynamic wave transmission reduces the energy flux to the crack tip. 

We identify the critical branching speed by matching the measured velocities with the planar versus branching classification of each crack. Branching first occurs at an initial stress of \SI{5.60}{\giga\pascal}, corresponding to a crack speed of \SI{2.4}{\kilo\meter\per\second}, or around $0.72\,v_R$. This is consistent with the highest crack velocities observed experimentally in fused silica \cite{mccauleyExperimentalObservationsDynamic2013}. In previous work with an empirical potential \cite{guren2022nanoscale}, the effective bond extensibility ($r_{\rm break}/r_0 = 1.144$) predicted a lower instability threshold, inconsistent with experimental branching speeds. The MLIP used here yields $r_{\rm break}/r_0 = 1.26$, which is consistent with a high instability threshold and in good agreement with crack instability theory for 2D brittle solids \cite{buehler2006dynamical}.

In the thin strip at steady state, $G(v) = \Gamma(v)$ \cite{finebergInstabilityDynamicFracture1999, bouchbinderDynamicsRapidFracture2014a}. Combining this with Eq.~\ref{eq:vc_sqrt} gives the velocity dependence of the fracture energy:
\begin{equation}
      \Gamma(v) = \frac{\Gamma_0}{1-(v/v_R)^2}.
      \label{eq:fracture_energy}
  \end{equation}
This represents a sharper transition from not cracking to fast cracking than predicted by the Freund model. In the absence of internal energy dissipation, linear elastic fracture mechanics in the thin strip predicts a step function from zero velocity to the Rayleigh speed, with an appreciable transient acceleration regime \cite{liuEnergySteadystateCrack1991}. Eq.~\ref{eq:vc_sqrt} yields a smoother transition than that, albeit sharper than the standard Freund model. This square-root form has consequences both for how fracture energy scales with velocity, which we examine next, and for the interpretation of crack-tip temperatures inferred from fractoluminescence experiments, which we return to below.
\subsection{Fracture energy}
The total fracture energy $\Gamma(v)=G(v)$ in steady state includes all dissipation at the crack tip. To disentangle whether the velocity-dependent fracture energy arises from increased surface area or from a change in the energy density of the surface, we define two post-mortem measures that isolate the energy stored in the damaged structure after the crack has passed and the material has cooled. The first is the structural fracture energy, obtained by dividing the stored structural energy by the projected crack area:
\begin{equation}
\Gamma_{\rm structural}=\frac{(E_{\rm p}^\text{after}-E_{\rm k}^\text{after}) - (E_{\rm p}^\text{before}-E_{\rm k}^\text{before})}{2A_{\rm proj}},
\label{eq:gamma_effective}
\end{equation}
where $E_p$ and $E_k$ are the total potential and kinetic energies of the system. The superscripts 'before' and 'after' denote the values before loading and after the crack has propagated. $A_{\rm proj}$ is the projected (nominal) area of the new crack, and the factor of 2 is due to a crack having two surfaces. The kinetic energy subtraction removes residual thermal energy, so that $\Gamma_\mathrm{structural}$ reflects only the structural cost of fracture. Since $\Gamma_\mathrm{structural}$ excludes heat, it is necessarily lower than the total fracture energy $\Gamma(v)$. The second measure is the intrinsic surface energy density, obtained by dividing the stored energy by the real (rough) surface area $A_{\rm real}$, measured using a Delaunay tessellation in Ovito \cite{stukowskiComputationalAnalysisMethods2014, stukowski2009visualization}, with a probe sphere radius of \SI{6}{\angstrom}, which is a reasonable choice to avoid measuring surfaces in the bulk of the material.
\begin{equation}
\gamma_s=\frac{(E_{\rm p}^\text{after}-E_{\rm k}^\text{after}) - (E_{\rm p}^\text{before}-E_{\rm k}^\text{before})}{A_{\rm real}}.
\label{eq:gamma_real}
\end{equation}
The ratio $\Gamma_\mathrm{structural}/\gamma_s = A_\mathrm{real}/(2A_\mathrm{proj})$ isolates the contribution of nanoscale roughening from the contribution of elevated surface energy density.

Fig.~\ref{fig:velocity}b shows how both $\Gamma_{\rm structural}$ and $\gamma_s$ increase with crack speed. Before branching occurs, $\Gamma_{\rm structural}$ rises by up to 33\%, while $\gamma_s$ increases by up to 15\%. The remaining increase is accounted for by a 16\% growth in the real crack area, $A_\text{real}$ (Fig.~\ref{fig:velocity}c). Even below the branching threshold, faster cracks create a larger volume of damaged material around the fracture surface, visible in Fig.~\ref{fig:morphology} as an increasingly thick damage zone. This is consistent with the growth of the process zone with stress intensity (Fig.~\ref{fig:setup}d), and shows that nanoscale roughening contributes to the fracture energy already at the atomic scale.

For the slowest crack that reaches the end of the simulation cell, we measure the post-mortem structural fracture energy to be $\Gamma_{\rm structural}=\SI{1.85}{\joule\per\meter\squared}$, and the local $\gamma_s=\SI{1.38}{\joule\per\meter\squared}$ for the real surface area. For comparison, the crack energy ($\Gamma$) in pure silica glass has been shown experimentally to be \SI{4.42}{\joule\per\metre\squared} \cite{wiederhorn1969fracture}, while molecular dynamics simulations of atomically flat silica surfaces estimates the surface energy ($\gamma$) to be \SI{1.33}{\joule\per\meter\squared} \cite{rimsza2018crack}. The close agreement between our slowest-crack $\gamma_s$ and the atomistically computed flat-surface energy \cite{rimsza2018crack} indicates that at low velocities, the newly created surfaces are structurally similar to relaxed equilibrium surfaces. 
\subsection{Heat and fractoluminescence}
A large proportion of the excess energy is dissipated as heat \cite{weichertHeatGenerationTip1978a}, producing localized temperature spikes high enough to drive blackbody photon emission, a phenomenon known as fractoluminescence \cite{pallares2012fractoluminescence,kadono2022experimental,zhang2022fracture} with spectra consistent with blackbody radiation \cite{pallares2012fractoluminescence}. \citet{pallares2012fractoluminescence} used the measured spectra together with the Freund velocity law (Eq.~\ref{eq:vc_classic}) to estimate crack speeds from crack tip temperatures. For example, they predicted that a crack tip temperature of $\sim$\SI{4980}{\kelvin} corresponds to a crack speed of $\sim$\SI{1.3}{\kilo\meter\per\second}.

We measure the crack tip temperatures directly in our simulations by averaging the kinetic energy per particle over \SI{0.5}{\pico\second} intervals, and converting to temperature via the equipartition theorem. \SI{0.5}{\pico\second} corresponds to roughly \SI{1}{\nano\meter} of resolution along the crack path of a fast crack running at \SI{2}{\kilo\meter\per\second}, allowing for spatial resolution of the temperature field within the crack process zone. An example of the per-particle temperature field during fast crack propagation is shown in Fig.~\ref{fig:setup}b. For a crack running at \SI{1.43}{\kilo\meter\per\second}, the temperature of the hottest 0.0001 percentile of particles in the system is \SI{3322}{\kelvin}. This temperature is substantially lower than the $\sim$\SI{4980}{\kelvin} predicted at comparable crack velocity by  \citet{pallares2012fractoluminescence}.

\begin{figure}
    \centering
    \includegraphics[width=\linewidth]{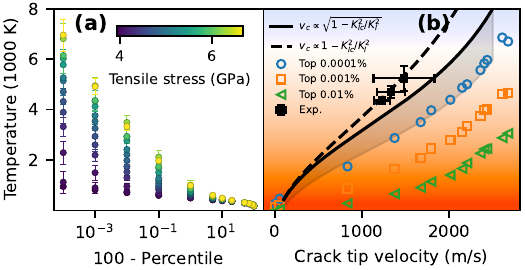}
    \caption{Thermal response and reconciliation with fractoluminescence. (a) Average kinetic temperature at selected atomic percentiles versus initial tensile stress. Error bars indicate fluctuations during steady-state propagation. (b) Crack tip temperature as a function of velocity. Open symbols show simulation data at selected atomic percentiles (the hottest 0.0001\%, 0.001\% and 0.01\% of atoms). The curves represent analytical predictions based on material properties (Table \ref{tbl:computed_values}): the dashed line shows the crack heat model by Pallares~et~al.~\cite{pallares2012fractoluminescence} (Eq.~\ref{eq:old_vc_dT_relationship}), while the solid line shows the law derived in the present study (Eq.~\ref{eq:new_vc_dT_relationship}).  The gray shaded area shows the effect of increasing Young's modulus from 68.9 up to \SI{90}{\giga\pascal}, to illustrate how much hyperelasticity may affect the model prediction. Filled black markers indicate experimental data from \citet{pallares2012fractoluminescence}: crack tip temperatures measured from fractoluminescence spectra, with velocities inferred using the Freund law (Eq.~\ref{eq:old_vc_dT_relationship}). The background gradient shows the black-body color at each temperature.}
    \label{fig:heat}
\end{figure}

This discrepancy is not due to the process zone size: our estimate of \SIrange{3}{6}{\nano\meter} is consistent with the $\sim$\SIrange{3}{4}{\nano\meter} range \citet{pallares2012fractoluminescence} assumed. Instead, it arises from the velocity--stress intensity relationship. Replacing the Freund law (Eq.~\ref{eq:vc_classic}) with the square-root relationship observed in our simulations (Eq.~\ref{eq:vc_sqrt}), while keeping the Dugdale--Barenblatt process zone and ordinary heat transport, the velocity--temperature relation changes. The relationship used in \citet{pallares2012fractoluminescence} is 
\begin{equation}
    v_c = v_R \frac{\alpha \Delta T^2}{K_{Ic}^2 + \alpha \Delta T^2},
     \label{eq:old_vc_dT_relationship}
 \end{equation}
 with $\alpha = E^2\rho ck/2v_R\sigma_y^2(1-\nu^2)^2$, where $\rho$, $c$, $k$ and $\nu$ are the density, specific heat capacity, thermal conductivity and Poisson's ratio of silica. $\sigma_y$ is the yield stress from the Dugdale--Barenblatt process zone formula. Using the square-root relationship (Eq.~\ref{eq:vc_sqrt}), we instead obtain  
 \begin{equation}
     v_c = v_R\left(\sqrt{1+\beta^2} - \beta\right),
     \label{eq:new_vc_dT_relationship}
 \end{equation}
with $\beta = K_{Ic}^2/(2\Delta T^2 \alpha)$. In Fig.~\ref{fig:heat}b, the filled markers from \citet{pallares2012fractoluminescence} lie on the dashed curve by construction: their crack tip temperatures were measured from fractoluminescence spectra, but the corresponding crack speeds were not measured directly---they were inferred using the Freund law (Eq.~\ref{eq:old_vc_dT_relationship}), which defines the dashed curve. In our simulations (open symbols), both temperature and velocity are measured independently. The apparent disagreement between the two datasets is largely resolved by the solid curve: replacing the Freund law with our square-root relationship (Eq.~\ref{eq:vc_sqrt}) shifts the predicted speed at a given temperature toward higher values, bringing the analytical prediction into closer agreement with the simulation data.
\section{Discussion and Conclusion}
The structural fracture energy in our simulations increases by up to 33\% below the branching threshold. \citet{sharonEnergyDissipationDynamic1996} observed a comparable increase in PMMA, which they attributed to an unidentified velocity-dependent dissipation process. Our simulations offer a possible explanation for this type of pre-branching increase in the fracture energy: in silica glass, we find that approximately half of the increase originates from nanoscale roughening that enlarges the real fracture surface area, while the remaining half is due to an elevated intrinsic surface energy density of the newly created surfaces. Since the nanoscale roughening occurs at length scales well below both typical macroscale instability patterns and the resolution of optical surface analysis, both contributions would appear as an increase in surface energy density or unresolved dissipation in a typical experiment (e.g. \cite{sharonMicrobranchingInstabilityDynamic1996}).

The elevated surface energy density is consistent with the extreme thermal conditions at the crack tip. At the velocities where $\gamma_s$ rises by 15\%, the crack tip reaches temperatures exceeding 3000\,K---well above the glass transition temperature of silica---within a process zone of only 3\,nm to 6\,nm. Our estimate of this zone bridges the gap between continuum models and atomistic theory: empirical potentials have historically overestimated its size ($\sim$10\,nm), whereas our \textit{ab initio} results align closely with the 3\,nm to 6\,nm scale following from the Dugdale--Barenblatt expression, and used in experimental fractoluminescence~\cite{pallares2012fractoluminescence}. The energy dissipated in this highly localized volume is sufficient to explain both the emission of light and the structurally altered, high-energy surfaces left in its wake.

A revised square-root velocity--stress intensity relationship (Eq.~\ref{eq:vc_sqrt}) further reconciles our atomistic crack tip temperatures with fractoluminescence measurements. Replacing the Freund law with this relationship shifts the predicted temperature at a given crack speed downward, explaining most of the discrepancy between our simulations and the predictions of \citet{pallares2012fractoluminescence}. 

The only physics assumption underlying these results is that the r$^2$SCAN functional faithfully represents the electronic structure of silica glass. With this assumption alone, we obtain close correspondence to experimental values for elastic properties, density, fracture toughness, and crack velocities---demonstrating that first-principles dynamic fracture simulations of fused silica can now achieve quantitative accuracy without empirical tuning.

In summary, our first-principles simulations show that the pre-branching fracture energy increase, first observed by \citet{sharonEnergyDissipationDynamic1996}, arises roughly equally from nanoscale roughening and an increase in the intrinsic surface energy density. These are contributions that standard post-mortem analysis would conflate. Our results demonstrate that first-principles molecular dynamics at realistic scales can now provide quantitative, experimentally validated insights into the process zone physics that governs dynamic fracture.
\section{Acknowledgements}
This work was supported by the Research Council of Norway through the Young Researcher Talent grants 344993 and 354100. We acknowledge the EuroHPC Joint Undertaking for awarding this project access to the EuroHPC supercomputer LUMI, hosted by CSC (Finland) and the LUMI consortium through a EuroHPC Regular Access call (Grants EHPC-REG-2023R02-088, EHPC-REG-2023R03-146). We thank Joachim Mathiesen for valuable discussions and feedback on an earlier draft of the manuscript.

Generative AI (Claude Opus 4.6--4.7, Anthropic) was used to assist with manuscript editing and revision. All authors have reviewed and take full responsibility for the content.
\section{End matter}
\subsection{Training of the Machine Learned Interatomic Potential (MLIP)}
We trained an E(3) equivariant MLIP, specifically an Allegro model \cite{musaelianLearningLocalEquivariant2023}, iteratively in an active learning loop to obtain an MLIP that was stable up to \SI{15000}{\kelvin}. The temperature range was motivated by the observation of fractoluminescence in silica glass fracture experiments \cite{pallares2012fractoluminescence}. The code and outputs from active learning, as well as the final model, are provided as Supplementary Data \cite{sveinsson_2025_17669954}, where the selected model that was used for fracture simulations is located in \lstinline{al_out_allegro_v4/iter_23/mlp/model_0}. 

The active learning loop was started from an initial database of 18 perturbed structures: six each of $\alpha$-quartz, $\beta$-cristobalite and coesite, with all the atomic positions perturbed by a stochastic amount drawn from a normal distribution with standard deviation of \SI{0.02}{\angstrom}, and forces and energies computed by Density Functional Theory (DFT), as described in the next section.  

Having an initial dataset, the active learning loop consists of the following steps:
\begin{itemize}
    \item Train 6 MLIPs with all collected data.
    \item Molecular dynamics sampling. We run simulations with various structures and conditions to generate a representative set of atomistic structures.
    \item Uncertainty-calibrated adversarial attacks \cite{cezarLearningAtomicForces2025} on randomly selected structures from the previous step. We drive the calibrated error estimate made using the MLIP models towards \SI{0.2}{\electronvolt\per\angstrom}.
    \item DFT calculations to label forces and energies on the structures from the previous step.
\end{itemize}
This procedure was run for a total of 24 iterations. The molecular dynamics sampling step changed throughout the active learning loop. During the first 12 iterations, all simulations were melt simulations of $\alpha$-quartz, $\beta$-cristobalite, and coesite. To span a wide pressure range, we scaled the simulations box lengths by $\{0.9, 0.95, 1.0, 1.05, 1.1\}$ from the equilibrium structures. In iterations 1--6, the temperature was gradually increased from 300 to \SI{7000}{\kelvin}, whereas in iterations 7--12, the temperature was instead increased to \SI{15000}{\kelvin}. For iterations 13--18, we moved on to melt-quenching, and simulated only $\beta$-cristobalite in a process of raising the temperature from \SI{300}{\kelvin} to \SI{6000}{\kelvin} over \SI{30}{\pico\second}, keeping the temperature at \SI{6000}{\kelvin} for \SI{30}{\pico\second}, cooling to \SI{4000}{\kelvin} over \SI{20}{\pico\second}, and then cooling again to \SI{300}{\kelvin} over \SI{370}{\pico\second}, i.e., a rate of \SI{1d13}{\kelvin\per\second}. The melt--quench simulations were run at pressures of $\{1, 100, 1000, 10000, 50000\}$ \si{bar}. Finally, for iterations 19--24 we ran mechanical deformation simulations at the same confining pressures, with temperatures of $\{300, 600, 900, 1200\}$~K, and conditions of compression to a strain of $\varepsilon=-0.3$, tension to $\varepsilon=1.0$ and simple shear at a boundary speed of \SI{10}{\meter\per\second}.

The MLIP was trained using Allegro~\cite{musaelianLearningLocalEquivariant2023}. We set a radial cutoff of $r_{\text{cut}} = \SI{5.5}{\angstrom}$ with a polynomial envelope function ($p=6$) and a radial basis of 8 trainable Bessel functions. We set the maximum angular momentum to $l_{\text{max}}=1$ with full O(3) parity and a single tensor product interaction layer. The network used 64 tensor features, while the two-body latent multilayer perceptron (MLP) and the scalar latent MLP were constructed with hidden dimensions of [128, 256, 256] and [256, 256, 256], respectively. The final edge energy MLP consisted of a single hidden layer with 128 units. All internal activation functions were SiLU. The model was trained using the Adam optimizer with an initial learning rate of $10^{-3}$ and a batch size of 2. The loss function was a weighted sum of contributions from total energy, forces, and stress, with weights of $\lambda_E=1.0$, $\lambda_F=1.0$, and $\lambda_S=10.0$, respectively. A learning rate scheduler (ReduceLROnPlateau) was used to reduce the learning rate by a factor of 0.5 if the validation loss did not improve for 50 epochs.

In the final iteration, we obtained a mean absolute error in the range of \SIrange{0.17}{0.2}{\electronvolt\per\angstrom} in the forces. This is higher than common when training MLIPs, but not necessarily high in a relative sense, since much of the training data is molten silica at temperatures up to \SI{7000}{\kelvin} and \SI{15000}{\kelvin}, and silica undergoing mechanical failure.

All DFT calculations were run using the Vienna Ab initio Simulation Package (VASP) \cite{kresseEfficiencyAbinitioTotal1996, kresseEfficientIterativeSchemes1996}, with the input settings shown in table \ref{tab:incar}. Since the target system that we need accurate forces and energies for (silica glass) is amorphous, and our systems contain 192 atoms and above, the Brillouin zone was sampled only at the $\Gamma$-point. We applied the standard PAW potential for silicon, and the hard PAW potential for oxygen. 
\begin{table}
\caption{VASP INCAR parameters.}
\centering
\begin{tabular}{ll}
\hline
Parameter & Value \\
\hline
ENCUT   & 1000 \\
EDIFF   & $1\times10^{-6}$ \\
ISYM    & 0 \\
PREC    & Accurate \\
ISMEAR  & 0 \\
SIGMA   & 0.1 \\
IBRION  & -1 \\
ALGO    & Normal \\
LREAL   & Auto \\
\hline
\end{tabular}
\label{tab:incar}
\end{table}
\subsection{Creating the amorphous silica glass}
The initial state for creating amorphous silica glass was a 192-atom cell of beta cristobalite replicated up to a super-cell of 1557504 atoms, with dimensions of $74.0\times8.54\times37.0$ \si{\nano\meter}$^3$. We applied the following melt-quench protocol using the Allegro MLIP in LAMMPS: The system was initialized at \SI{300}{\kelvin} and heated to \SI{5500}{\kelvin} over \SI{30}{\pico\second} in the NVT ensemble, followed by equilibration at \SI{5500}{\kelvin} for \SI{100}{\pico\second} to ensure a fully melted state. The melt was then cooled to \SI{4000}{\kelvin} over \SI{150}{\pico\second} in the NVT ensemble. 

From \SI{4000}{\kelvin}, the system was quenched to \SI{300}{\kelvin} in the NPT ensemble at a cooling rate of \SI{1}{\kelvin\per\pico\second} ($\SI{e12}{\kelvin\per\second}$) while maintaining zero external pressure (\SI{1}{\bar}). Finally, the quenched glass was equilibrated at \SI{300}{\kelvin} and \SI{1}{\bar} for \SI{100}{\pico\second} in the NPT ensemble. Temperature and pressure were controlled using a Nosé-Hoover thermostat and barostat with damping constants of \SI{0.1}{\pico\second} and \SI{1.0}{\pico\second}, respectively. Anisotropic cell deformations were allowed during the final equilibration phase to relax residual stresses. This system was replicated 4 times in the $x$ and 2 times in the $z$ direction to create the production system of 12.3 million atoms (after carving the notch). 
\subsection{Elastic parameters}
\begin{figure}
    \centering
    \includegraphics{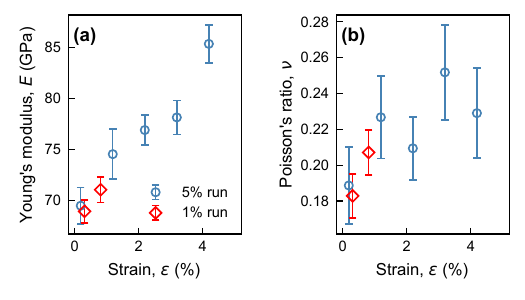}
    \caption{Calculation of Young's modulus $E$ and Poisson's ratio $\nu$ for our model silica glass. (a) $E$ versus applied strain. (b) $\nu$ versus applied strain. Two independent runs are shown (1\% and 5\% maximum strain). Error bars are bootstrap estimates.}
    \label{fig:E_nu}
\end{figure}

To obtain the elastic parameters of amorphous silica created with the MLIP, we ran a separate melt-quench simulation of a system of 12288 atoms, measuring around $5.7\times5.7\times\SI{5.7}{\nano\meter}^3$, with the same protocol as described in the previous paragraph. We relaxed the system and measured the density to $\SI{2.17}{\gram\per\cm\cubed}$. Then, we ran two simulations where we gradually imposed a strain in the $x$ direction, while the $y$ and $z$ directions were kept at a constant pressure of \SI{1}{\bar}. For the first simulation we imposed 1 \% strain over \SI{50}{\pico\second} and for the second one we imposed 5 \% strain over \SI{100}{\pico\second}. The  Young's modulus was obtained by linear fits of the stress-strain curve, and the Poisson's ratio by linear fits of the lateral versus axial strains. Values obtained at various strain levels are shown in Fig. \ref{fig:E_nu}. The uncertainties were computed by bootstrap resampling. The values obtained closest to equilibrium, at 0.25 \% strain were a Young's modulus of $E=\SI{68.9\pm1.2}{\giga\pascal}$ and a Poisson's ratio of $\nu=0.183\pm0.012$, both of which are within 5 \% of typical experimental values of $E\approx\SI{72}{\giga\pascal}$ and $\nu\approx0.16–0.176$ (e.g. \cite{pallares2012fractoluminescence, deschampsElasticModuliPermanently2014}). The Young's modulus and Poisson's ratios are somewhat sensitive to the strain, as can be seen from Fig. \ref{fig:E_nu}. 
\subsection{Determining the size of the process zone}
\begin{figure}
    \centering
    \includegraphics[width=\linewidth]{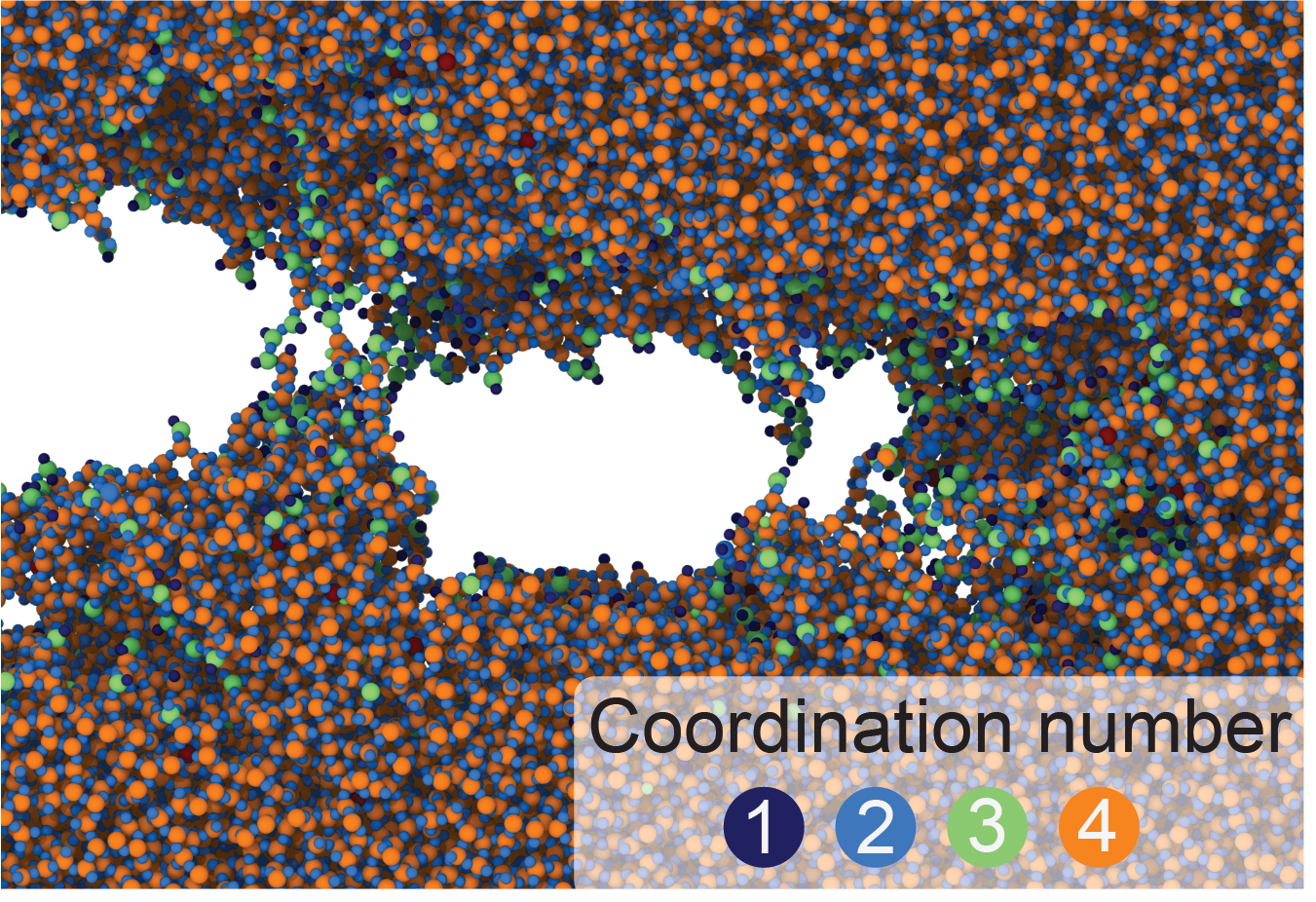}
    \caption{Snapshot of the crack tip where the atoms are colored after the coordination. The under-coordinated atoms are used to find the position of the crack tip and the size of the process zone.}
    \label{fig:coordination}
\end{figure}

Ahead of the crack tip, there is a region with under-coordinated atoms (Fig. \ref{fig:coordination}), and the radius of the process zone ($r_{pz}$) was located by finding the largest cluster among these under-coordinated atoms. Fig.~\ref{fig:setup}c shows the two largest clusters, colored pink and green, where the green cluster is the largest and ahead of the crack tip. The radius was calculated using $r_{pz}=2\sqrt{\lambda_{min}}$, where $\lambda_{min}$ is the shortest length of cylinder that expresses the shape of the process zone.
\begin{table}
    \centering
    \caption{Values used for computing the relationship between crack speeds and crack tip temperatures, using Eqs. \ref{eq:old_vc_dT_relationship} and \ref{eq:new_vc_dT_relationship}.}
    \label{tbl:computed_values}
    \begin{tabular}{ccc}
    \hline
    \textbf{Property} & \textbf{Value} & \textbf{Source} \\
    \hline
    Young's modulus $E$ (GPa) & 68.9 & Present study \\
    Poisson's ratio $\nu$ & 0.183 & Present study \\
    Density $\rho$ (g/cm$^3$) & 2.17 & Present study \\
    Fracture toughness $K_{IC}$($\si{\mega\pascal\sqrt{\meter}}$) & 0.763 & Present study\\
    Ultimate strength of glass $\sigma_Y$ (GPa) & 10 & Literature \cite{pallares2012fractoluminescence} \\
    Specific heat capacity $c$ (\si{\joule\per\kilo\gram\per\kelvin}) & 703 & Literature \cite{pallares2012fractoluminescence} \\ 
    Heat conductivity $k$ (\si{\watt\per\meter\per\kelvin})& 1.30 & Literature \cite{pallares2012fractoluminescence} \\ 
    \hline
    \end{tabular}
\end{table}
\bibliography{bibliography}

@book{buehlerAtomisticModelingMaterials2008,
  title = {Atomistic {{Modeling}} of {{Materials Failure}}},
  editor = {Buehler, Markus J.},
  year = {2008},
  publisher = {Springer US},
  address = {Boston, MA},
  doi = {10.1007/978-0-387-76426-9},
  urldate = {2022-04-21},
  isbn = {978-0-387-76425-2 978-0-387-76426-9},
  langid = {english},
}

@dataset{sveinsson_2025_17669954,
  author       = {Sveinsson, Henrik Andersen},
  title        = {Allegro machine learned interatomic potential for silica up to 15000 K},
  month        = nov,
  year         = 2025,
  publisher    = {Zenodo},
  doi          = {10.5281/zenodo.17669954},
  url          = {https://doi.org/10.5281/zenodo.17669954},
}

@article{cezarLearningAtomicForces2025,
  title = {Learning Atomic Forces from Uncertainty-Calibrated Adversarial Attacks},
  author = {Cezar, Henrique Musseli and Bodenstein, Tilmann and Sveinsson, Henrik Andersen and Ledum, Morten and Reine, Simen and Bore, Sigbj{\o}rn L{\o}land},
  year = 2025,
  month = jul,
  journal = {npj Computational Materials},
  volume = {11},
  number = {1},
  pages = {200},
  issn = {2057-3960},
  doi = {10.1038/s41524-025-01703-5},
  urldate = {2025-10-06},
  langid = {english},
}

@article{rimsza2018crack,
  title={Crack propagation in silica from reactive classical molecular dynamics simulations},
  author={Rimsza, Jessica M and Jones, Reese E and Criscenti, Louise J},
  journal={Journal of the American Ceramic Society},
  volume={101},
  number={4},
  pages={1488--1499},
  year={2018},
  publisher={Wiley Online Library}
}

@article{rountree2010fracture,
  title={Fracture in glass via molecular dynamics simulations and atomic force microscopy experiments},
  author={Rountree, CL and Bonamy, D and Dalmas, D and Prades, S and Kalia, RK and Guillot, C and Bouchaud, E},
  journal={Physics and Chemistry of Glasses-European Journal of Glass Science and Technology Part B},
  volume={51},
  number={2},
  pages={127--132},
  year={2010},
  publisher={Society of Glass Technology}
}

@article{balyakin2020deep,
  title={Deep machine learning interatomic potential for liquid silica},
  author={Balyakin, IA and Rempel, SV and Ryltsev, RE and Rempel, AA},
  journal={Physical Review E},
  volume={102},
  number={5},
  pages={052125},
  year={2020},
  publisher={APS}
}

@article{behlerGeneralizedNeuralNetworkRepresentation2007,
  title = {Generalized {{Neural-Network Representation}} of {{High-Dimensional Potential-Energy Surfaces}}},
  author = {Behler, J{\"o}rg and Parrinello, Michele},
  year = 2007,
  month = apr,
  journal = {Physical Review Letters},
  volume = {98},
  number = {14},
  pages = {146401},
  issn = {0031-9007, 1079-7114},
  doi = {10.1103/PhysRevLett.98.146401},
  urldate = {2025-10-22},
  copyright = {http://link.aps.org/licenses/aps-default-license},
  langid = {english},
}

@article{vashishtaInteractionPotentialSiO21990,
  title = {Interaction Potential for {{SiO2}} : {{A}} Molecular-Dynamics Study of Structural Correlations},
  shorttitle = {Interaction Potential for {{SiO}} 2},
  author = {Vashishta, P. and Kalia, Rajiv K. and Rino, Jos{\'e} P. and Ebbsj{\"o}, Ingvar},
  year = 1990,
  month = jun,
  journal = {Physical Review B},
  volume = {41},
  number = {17},
  pages = {12197--12209},
  issn = {0163-1829, 1095-3795},
  doi = {10.1103/PhysRevB.41.12197},
  urldate = {2022-04-21},
  langid = {english},
}

@article{novikov2019improving,
  title={Improving accuracy of interatomic potentials: more physics or more data? A case study of silica},
  author={Novikov, Ivan S and Shapeev, Alexander V},
  journal={Materials Today Communications},
  volume={18},
  pages={74--80},
  year={2019},
  publisher={Elsevier}
}

@article{rountree2020silica,
  title={SILICA and its process zone},
  author={Rountree, Cindy L and Feng, Weiying},
  journal={International Journal of Applied Glass Science},
  volume={11},
  number={3},
  pages={385--395},
  year={2020},
  publisher={Wiley Online Library}
}

@article{weichertHeatGenerationTip1978a,
  title = {Heat Generation at the Tip of a Moving Crack},
  author = {Weichert, R. and Sch{\"o}nert, K.},
  year = {1978},
  month = jun,
  journal = {Journal of the Mechanics and Physics of Solids},
  volume = {26},
  number = {3},
  pages = {151--161},
  issn = {00225096},
  doi = {10.1016/0022-5096(78)90006-6},
  urldate = {2025-10-13},
  copyright = {https://www.elsevier.com/tdm/userlicense/1.0/},
  langid = {english},
}

@article{zengDeePMDkitV2Software2023,
  title = {{{DeePMD-kit}} v2: {{A}} Software Package for Deep Potential Models},
  shorttitle = {{{DeePMD-kit}} V2},
  author = {Zeng, Jinzhe and Zhang, Duo and Lu, Denghui and Mo, Pinghui and Li, Zeyu and Chen, Yixiao and Rynik, Mari{\'a}n and Huang, Li'ang and Li, Ziyao and Shi, Shaochen and Wang, Yingze and Ye, Haotian and Tuo, Ping and Yang, Jiabin and Ding, Ye and Li, Yifan and Tisi, Davide and Zeng, Qiyu and Bao, Han and Xia, Yu and Huang, Jiameng and Muraoka, Koki and Wang, Yibo and Chang, Junhan and Yuan, Fengbo and Bore, Sigbj{\o}rn L{\o}land and Cai, Chun and Lin, Yinnian and Wang, Bo and Xu, Jiayan and Zhu, Jia-Xin and Luo, Chenxing and Zhang, Yuzhi and Goodall, Rhys E. A. and Liang, Wenshuo and Singh, Anurag Kumar and Yao, Sikai and Zhang, Jingchao and Wentzcovitch, Renata and Han, Jiequn and Liu, Jie and Jia, Weile and York, Darrin M. and E, Weinan and Car, Roberto and Zhang, Linfeng and Wang, Han},
  year = 2023,
  month = aug,
  journal = {The Journal of Chemical Physics},
  volume = {159},
  number = {5},
  pages = {054801},
  issn = {0021-9606, 1089-7690},
  doi = {10.1063/5.0155600},
  urldate = {2023-11-03},
  langid = {english},
}

@article{stukowskiComputationalAnalysisMethods2014,
  title = {Computational {{Analysis Methods}} in {{Atomistic Modeling}} of {{Crystals}}},
  author = {Stukowski, Alexander},
  year = 2014,
  month = mar,
  journal = {JOM},
  volume = {66},
  number = {3},
  pages = {399--407},
  issn = {1047-4838, 1543-1851},
  doi = {10.1007/s11837-013-0827-5},
  urldate = {2025-11-28},
  copyright = {http://www.springer.com/tdm},
  langid = {english},
}

@article{liuEnergySteadystateCrack1991,
  title = {The Energy of a Steady-State Crack in a Strip},
  author = {Liu, Xiangming and Marder, M.},
  year = 1991,
  month = jan,
  journal = {Journal of the Mechanics and Physics of Solids},
  volume = {39},
  number = {7},
  pages = {947--961},
  issn = {00225096},
  doi = {10.1016/0022-5096(91)90013-E},
  urldate = {2025-12-12},
  copyright = {https://www.elsevier.com/tdm/userlicense/1.0/},
  langid = {english},
}

@article{bouchbinderDynamicsRapidFracture2014a,
  title = {The Dynamics of Rapid Fracture: Instabilities, Nonlinearities and Length Scales},
  shorttitle = {The Dynamics of Rapid Fracture},
  author = {Bouchbinder, Eran and Goldman, Tamar and Fineberg, Jay},
  year = 2014,
  month = apr,
  journal = {Reports on Progress in Physics},
  volume = {77},
  number = {4},
  pages = {046501},
  issn = {0034-4885, 1361-6633},
  doi = {10.1088/0034-4885/77/4/046501},
  urldate = {2025-12-09},
  copyright = {http://iopscience.iop.org/info/page/text-and-data-mining},
  langid = {english},
}

@article{finebergInstabilityDynamicFracture1999,
  title={Instability in dynamic fracture},
  author={Fineberg, Jay and Marder, M},
  journal={Physics Reports},
  volume={313},
  number={1-2},
  pages={1--108},
  year={1999},
  publisher={Elsevier}
}

@article{sharonMicrobranchingInstabilityDynamic1996,
  title = {Microbranching Instability and the Dynamic Fracture of Brittle Materials},
  author = {Sharon, Eran and Fineberg, Jay},
  year = 1996,
  month = sep,
  journal = {Physical Review B},
  volume = {54},
  number = {10},
  pages = {7128--7139},
  issn = {0163-1829, 1095-3795},
  doi = {10.1103/PhysRevB.54.7128},
  urldate = {2026-03-26},
  copyright = {http://link.aps.org/licenses/aps-default-license},
  langid = {english},
}

@article{stukowski2009visualization,
  title={Visualization and analysis of atomistic simulation data with OVITO--the Open Visualization Tool},
  author={Stukowski, Alexander},
  journal={Modelling and simulation in materials science and engineering},
  volume={18},
  number={1},
  pages={015012},
  year={2009},
  publisher={IOP Publishing}
}

@book{freund1998dynamic,
  title={Dynamic fracture mechanics},
  author={Freund, Lambert Ben},
  year={1998},
  publisher={Cambridge university press}
}

@inproceedings{NEURIPS2022_4a36c3c5,
 author = {Batatia, Ilyes and Kovacs, David P and Simm, Gregor and Ortner, Christoph and Csanyi, Gabor},
 booktitle = {Advances in Neural Information Processing Systems},
 editor = {S. Koyejo and S. Mohamed and A. Agarwal and D. Belgrave and K. Cho and A. Oh},
 pages = {11423--11436},
 publisher = {Curran Associates, Inc.},
 title = {MACE: Higher Order Equivariant Message Passing Neural Networks for Fast and Accurate Force Fields},
 url = {https://proceedings.neurips.cc/paper_files/paper/2022/file/4a36c3c51af11ed9f34615b81edb5bbc-Paper-Conference.pdf},
 volume = {35},
 year = {2022}
}

@article{sharonEnergyDissipationDynamic1996,
  title = {Energy {{Dissipation}} in {{Dynamic Fracture}}},
  author = {Sharon, Eran and Gross, Steven P. and Fineberg, Jay},
  year = 1996,
  month = mar,
  journal = {Physical Review Letters},
  volume = {76},
  number = {12},
  pages = {2117--2120},
  issn = {0031-9007, 1079-7114},
  doi = {10.1103/PhysRevLett.76.2117},
  urldate = {2025-12-09},
  copyright = {http://link.aps.org/licenses/aps-default-license},
  langid = {english},
}

@article{finebergInstabilityDynamicFracture1991,
  title = {Instability in Dynamic Fracture},
  author = {Fineberg, Jay and Gross, Steven P. and Marder, M. and Swinney, Harry L.},
  year = 1991,
  month = jul,
  journal = {Physical Review Letters},
  volume = {67},
  number = {4},
  pages = {457--460},
  issn = {0031-9007},
  doi = {10.1103/PhysRevLett.67.457},
  urldate = {2022-04-28},
  langid = {english},
}

@article{erhardMachinelearnedInteratomicPotential2022,
  title = {A Machine-Learned Interatomic Potential for Silica and Its Relation to Empirical Models},
  author = {Erhard, Linus C. and Rohrer, Jochen and Albe, Karsten and Deringer, Volker L.},
  year = {2022},
  month = dec,
  journal = {npj Computational Materials},
  volume = {8},
  number = {1},
  pages = {90},
  issn = {2057-3960},
  doi = {10.1038/s41524-022-00768-w},
  urldate = {2022-10-05},
  langid = {english},
}

@article{deschampsElasticModuliPermanently2014,
  title = {Elastic {{Moduli}} of {{Permanently Densified Silica Glasses}}},
  author = {Deschamps, T. and Margueritat, J. and Martinet, C. and Mermet, A. and Champagnon, B.},
  year = 2014,
  month = nov,
  journal = {Scientific Reports},
  volume = {4},
  number = {1},
  pages = {7193},
  issn = {2045-2322},
  doi = {10.1038/srep07193},
  urldate = {2026-04-19},
  langid = {english},
}

@article{kresseEfficiencyAbinitioTotal1996,
  title = {Efficiency of Ab-Initio Total Energy Calculations for Metals and Semiconductors Using a Plane-Wave Basis Set},
  author = {Kresse, G. and Furthm{\"u}ller, J.},
  year = {1996},
  month = jul,
  journal = {Computational Materials Science},
  volume = {6},
  number = {1},
  pages = {15--50},
  issn = {09270256},
  doi = {10.1016/0927-0256(96)00008-0},
  urldate = {2025-09-24},
  copyright = {https://www.elsevier.com/tdm/userlicense/1.0/},
  langid = {english},
}

@article{kresseEfficientIterativeSchemes1996,
  title = {Efficient Iterative Schemes for {\emph{Ab Initio}} Total-Energy Calculations Using a Plane-Wave Basis Set},
  author = {Kresse, G. and Furthm{\"u}ller, J.},
  year = {1996},
  month = oct,
  journal = {Physical Review B},
  volume = {54},
  number = {16},
  pages = {11169--11186},
  issn = {0163-1829, 1095-3795},
  doi = {10.1103/PhysRevB.54.11169},
  urldate = {2025-09-24},
  copyright = {http://link.aps.org/licenses/aps-default-license},
  langid = {english},
}

@article{musaelianLearningLocalEquivariant2023,
  title = {Learning Local Equivariant Representations for Large-Scale Atomistic Dynamics},
  author = {Musaelian, Albert and Batzner, Simon and Johansson, Anders and Sun, Lixin and Owen, Cameron J. and Kornbluth, Mordechai and Kozinsky, Boris},
  year = {2023},
  month = feb,
  journal = {Nature Communications},
  volume = {14},
  number = {1},
  pages = {579},
  issn = {2041-1723},
  doi = {10.1038/s41467-023-36329-y},
  urldate = {2024-05-10},
  langid = {english},
}

@article{furnessAccurateNumericallyEfficient2020,
  title = {Accurate and {{Numerically Efficient}} R{\textsuperscript{2}} {{SCAN Meta-Generalized Gradient Approximation}}},
  author = {Furness, James W. and Kaplan, Aaron D. and Ning, Jinliang and Perdew, John P. and Sun, Jianwei},
  year = {2020},
  month = oct,
  journal = {The Journal of Physical Chemistry Letters},
  volume = {11},
  number = {19},
  pages = {8208--8215},
  issn = {1948-7185, 1948-7185},
  doi = {10.1021/acs.jpclett.0c02405},
  urldate = {2025-09-24},
  copyright = {https://pubs.acs.org/page/policy/authorchoice\_termsofuse.html},
  langid = {english},
}

@article{kobayashiMachineLearningMolecular2023,
  title = {Machine Learning Molecular Dynamics Reveals the Structural Origin of the First Sharp Diffraction Peak in High-Density Silica Glasses},
  author = {Kobayashi, Keita and Okumura, Masahiko and Nakamura, Hiroki and Itakura, Mitsuhiro and Machida, Masahiko and Urata, Shingo and Suzuya, Kentaro},
  year = {2023},
  month = nov,
  journal = {Scientific Reports},
  volume = {13},
  number = {1},
  pages = {18721},
  issn = {2045-2322},
  doi = {10.1038/s41598-023-44732-0},
  urldate = {2025-01-10},
  langid = {english},
}

@article{guren2022nanoscale,
  title={Nanoscale damage production by dynamic tensile rupture in $\alpha$-quartz},
  author={Guren, Marthe G and Sveinsson, Henrik A and Malthe-S{\o}renssen, Anders and Renard, Fran{\c{c}}ois},
  journal={Geophysical Research Letters},
  volume={49},
  number={20},
  pages={e2022GL100468},
  year={2022},
  publisher={Wiley Online Library}
}

@article{wiederhorn1969fracture,
  title={Fracture surface energy of glass},
  author={Wiederhorn, So M},
  journal={Journal of the American Ceramic Society},
  volume={52},
  number={2},
  pages={99--105},
  year={1969},
  publisher={Wiley Online Library}
}

@article{buehler2006dynamical,
  title={Dynamical fracture instabilities due to local hyperelasticity at crack tips},
  author={Buehler, Markus J and Gao, Huajian},
  journal={Nature},
  volume={439},
  number={7074},
  pages={307--310},
  year={2006},
  publisher={Nature Publishing Group UK London}
}

@article{kadono2022experimental,
  title={Experimental Investigation of Visible-Light and X-ray Emissions during Rock and Mineral Fracture: Role of Electrons Traveling between Fracture Surfaces},
  author={Kadono, Toshihiko and Ogawa, Kazunori and Shirai, Kei and Arakawa, Masahiko and Kurosawa, Kosuke and Okamoto, Takaya and Matsui, Takafumi and Hasegawa, Sunao and Suzuki, Ayako I and Kobayashi, Hideyuki},
  journal={Minerals},
  volume={12},
  number={6},
  pages={778},
  year={2022},
  publisher={MDPI}
}

@article{pallares2012fractoluminescence,
  title={Fractoluminescence characterization of the energy dissipated during fast fracture of glass},
  author={Pallares, Gael and Rountree, Cindy L and Douillard, Ludovic and Charra, Fabrice and Bouchaud, Elisabeth},
  journal={Europhysics Letters},
  volume={99},
  number={2},
  pages={28003},
  year={2012},
  publisher={IOP Publishing}
}

@article{zhang2022fracture,
  title={Fracture of silicate glasses: Microcavities and correlations between atomic-level properties},
  author={Zhang, Zhen and Ispas, Simona and Kob, Walter},
  journal={Physical Review Materials},
  volume={6},
  number={8},
  pages={085601},
  year={2022},
  publisher={APS}
}

@article{mccauleyExperimentalObservationsDynamic2013,
  title = {Experimental {{Observations}} on {{Dynamic Response}} of {{Selected Transparent Armor Materials}}},
  author = {McCauley, J. W. and Strassburger, E. and Patel, P. and Paliwal, B. and Ramesh, K. T.},
  year = {2013},
  month = jan,
  journal = {Experimental Mechanics},
  volume = {53},
  number = {1},
  pages = {3--29},
  issn = {0014-4851, 1741-2765},
  doi = {10.1007/s11340-012-9658-5},
  urldate = {2025-09-23},
  copyright = {http://www.springer.com/tdm},
  langid = {english},
}

@article{dugdale1960yielding,
  title={Yielding of steel sheets containing slits},
  author={Dugdale, Donald S},
  journal={Journal of the Mechanics and Physics of Solids},
  volume={8},
  number={2},
  pages={100--104},
  year={1960},
  publisher={Elsevier}
}

@article{barenblatt1962mathematical,
  title={The mathematical theory of equilibrium cracks in brittle fracture},
  author={Barenblatt, Grigory Isaakovich},
  journal={Advances in applied mechanics},
  volume={7},
  pages={55--129},
  year={1962},
  publisher={Elsevier}
}
\end{document}